\documentclass[12pt]{article}

\usepackage{tikz}
\usepackage{amsmath}

\usepackage{makeidx,amssymb,amsmath,amsthm,graphicx}

\pdfoutput=1

\begin{document}

\title{Characterization of \\ YbNi$_4$(P$_{1-{\it x}}$As$_{\it x}$)$_2$, $x = 0, 0.2$ \\ single crystals grown by Czochralski method}
\author{K ~Kliemt and C ~Krellner}
\date{Institute of Physics, Goethe-University Frankfurt, Max-von-Laue Stasse 1, 60438 Frankfurt, Germany\\
kliemt@physik.uni-frankfurt.de}
\maketitle
\begin{abstract}
We have investigated large single crystals of YbNi$_4$P$_2$ that were grown 
from a levitating melt by the Czochralski method.
The new samples facilitate the determination of the absolute values of the electrical resistivity. Phase pure polycrystalline samples of the non-magnetic reference LuNi$_4$P$_2$ were prepared and the electrical resistivity was measured. 
Furthermore we have grown a single crystal of the As substituted compound YbNi$_4$(P$_{1-{\it x}}$As$_{\it x}$)$_2$, $x = 0.2$ and investigated the homogenity of the As distribution.

\end{abstract}


\section{Introduction}
YbNi$_4$P$_2$ has a Curie temperature of T$_{\rm C}$= 0.17\, K \cite{Krellner2011}. 
This is one of the lowest Curie temperatures 
ever observed among stoichiometric compounds \cite{Brando2016}. 
The Curie temperature can be decreased even further
in the substitution series YbNi$_4$(P$_{1-{\it x}}$As$_{\it x}$)$_2$.
At ${\it x}\approx$ 0.1 the rare case of ferromagnetic
quantum critical point (FM QCP) occurs \cite{Steppke2013}. 
YbNi$_4$P$_2$ was  found to be a Kondo lattice system 
with a Kondo temperature
of T$_{\rm K}\approx 8$\,K revealed from specific heat measurements on powder samples. 
The occurence of coherent Kondo scattering 
was observed in the resistivity
which steeply decreases below 20\,K \cite{Krellner2011}. 
Resistivity measurements from 2 - 300\,K on small Bridgman grown YbNi$_4$P$_2$
single crystals show an anisotropy between
the $j\parallel c$ and the $j\perp c$-direction,
which was explained by the presence of the
crystalline electric field \cite{Krellner2012}.
A further measurement on single crystals
gave absolute values at 300\,K 
of  $90\,\mu\Omega \rm cm$ for  $j\parallel c$
and  $160\,\mu\Omega \rm cm$ for $j\perp c$-direction
\cite{Steppke2013}.
Until now, all reported resistivity measurements on single crystals 
were performed on crystals grown by the Bridgman method.
This method yields thin needle shaped crystals 
of about $2-3\,$mm along the $[001]-$direction and  
of $0.5-0.7\,$mm in the perpendicular extension. 

Here, we have investigated the electrical transport properties 
down to 1.8\,K 
of single crystal samples 
which were cut from one large crystal.
This single crystal had a conical shape with a diameter 
of $3 - 9\,$mm and a length of 14 mm. It belongs to a new
generation of single crystals 
grown from a levitating melt by the Czochralski method  \cite{Kliemt2016}. 

In the past, samples of the substitution series 
YbNi$_4$(P$_{1-{\it x}}$As$_{\it x}$)$_2$ with 
As-contents up to $x=0.13$ \cite{Steppke2013} 
have been investigated and the material was studied 
close to the ferromagnetic QCP. 
Until now, the T-$x$ phase diagram is incomplete 
due to a lack of samples with $x>0.13$. 
To observe the emergence of Fermi liquid behaviour in the system,
samples with these higher As concentrations are required. 
Using a seed from the new crystal generation,
we have grown a single crystal of the substituted compound
YbNi$_4$(P$_{1-{\it x}}$As$_{\it x}$)$_2$ with $x = 0.2$
with 16\,mm in length and $3-6\,$mm in diameter.
It turned out that the As distribution is homogeneous.

\section{Experimental}
An YbNi$_4$P$_2$ single crystal was grown 
by the Czochralski method from a Ni-P self-flux 
as described in \cite{Kliemt2016}. The crystal structure was confirmed 
by powder X-ray diffraction (Cu K$_{\alpha}$). 
The chemical composition was checked 
by energy-dispersive X-ray spectroscopy (EDX). 
Upon investigating a polished surface 
by scanning electron microscopy and polarization microscopy 
an excellent homogenity of this surface 
and the absence of Ni-P flux-inclusions was found. 
The single crystal was oriented using the
X-ray Laue back-scattering technique. 
From this crystal, samples for the electrical transport
measurement were cut using a spark erosion device. 
The samples had initial dimensions of 
$\approx$ 3 mm x 0.5 mm x 0.15 mm and were oriented along 
the crystallographic main symmetry directions $[100]$, $[110]$ and $[001]$.
Electrical transport measurements were performed between 
1.8 and 300 $\rm K$ using the ACT option of a commercial PPMS 
(Quantum Design).
The Czochralski method was also applied to grow a single crystal
from a melt where 20 at\% of P was replaced by As. 
The crystal structure and the chemical composition were investigated 
by powder X-ray diffraction (PXRD) and EDX, respectively. 
Phase pure polycrystalline samples 
of the non-magnetic reference compound LuNi$_4$P$_2$ were prepared 
according to \cite{Chikhrij1986, Jeitschko1990}. 
PXRD measurements 
confirmed the $P4_2/mnm$ tetragonal structure with lattice parameters 
a = 7.047(6) \AA\,  and c = 3.583(4) \AA, which is in agreement with the 
data published for polycrystalline samples \cite{Chikhrij1986}. 
PXRD data were refined using 
the General Structure Analysis System (GSAS) \cite{GSAS, EXPGUI}.


\section{Electrical resistivity $\rho(T)$ of YbNi$_4$P$_2$}

\begin{figure}[ht]
\centering
\includegraphics[width=0.49\textwidth]{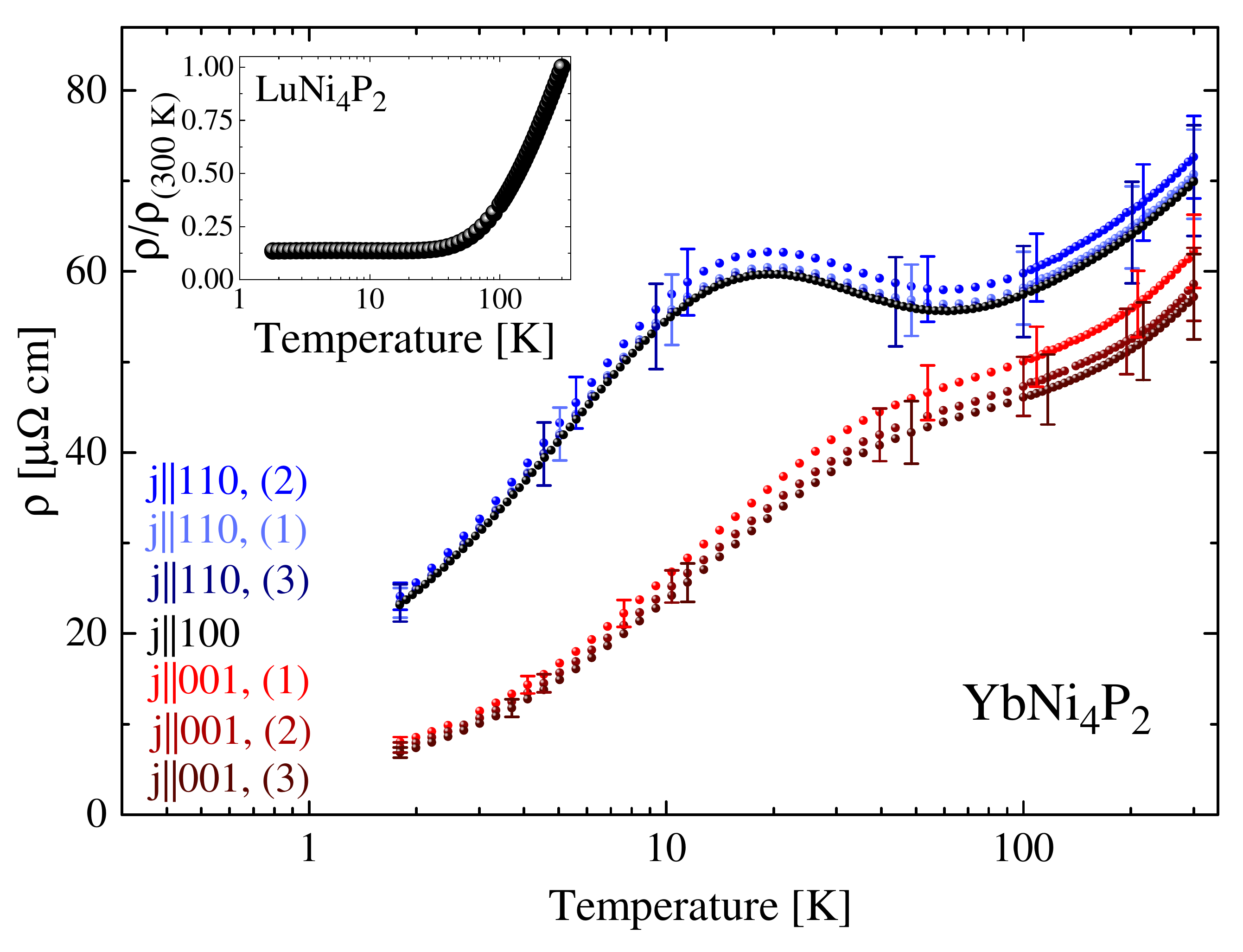}
\includegraphics[width=0.49\textwidth]{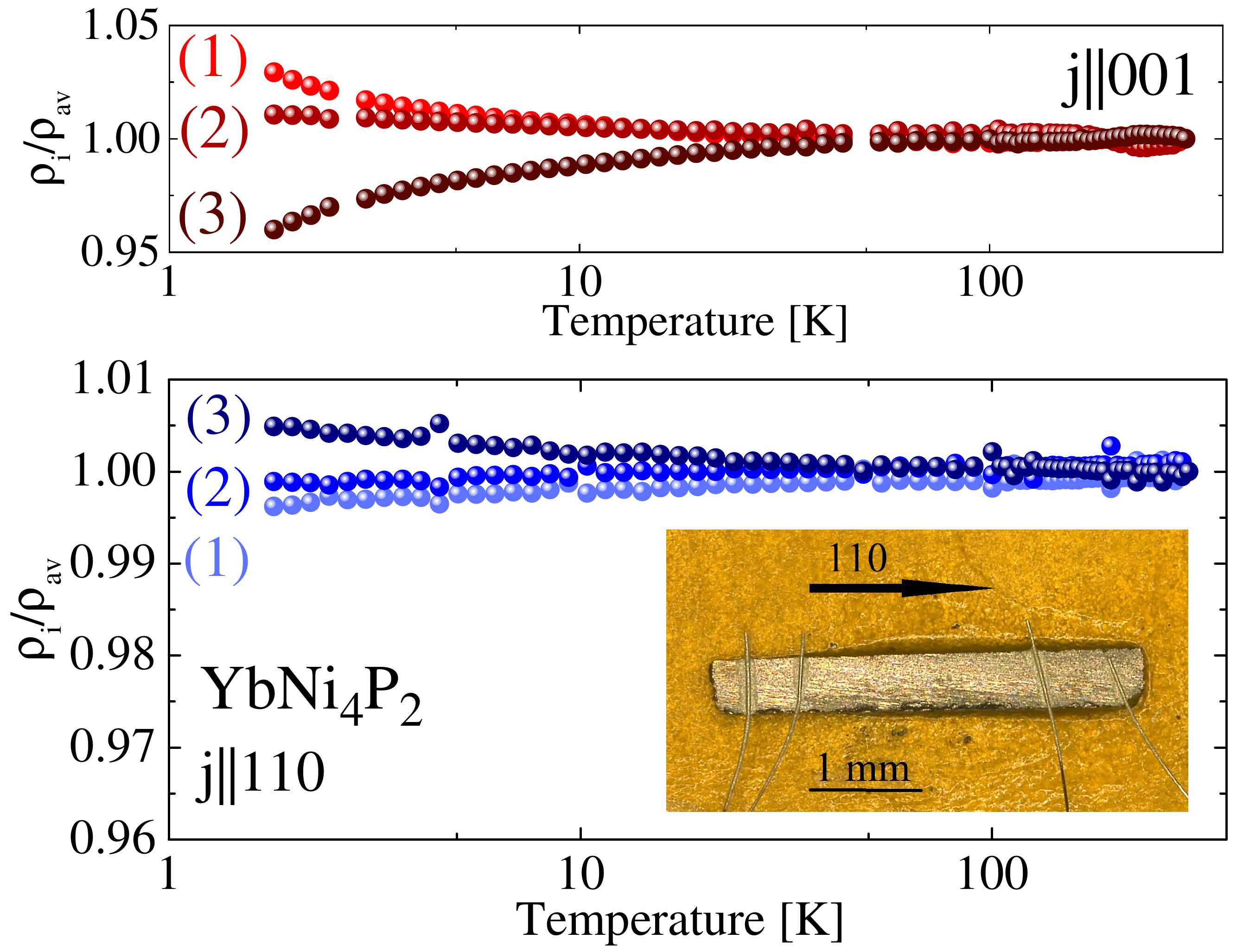}
\caption{{\it Left:} Electrical resistivity $\rho$ as a function of temperature from $1.8 \rm K$ to $ 300 \rm K$ measured with current $j$ perpendicular ($j\parallel 100$ and $j\parallel 110$) and parallel to the crystallographic c-axis.  The inset shows the electrical resistivity measured on a polycrystalline sample of LuNi$_4$P$_2$. The resistivity ratio was determined to be $\rho_{\rm poly}(300 \rm K)/\rho_{ \rm poly}(1.8\rm K)=7.4 $. {\it Right:} Ratio of the normalized electrical resistivity $\rho_i$ of measurement $i$ divided by the average resitivity $\rho_{\rm av}$ for the respective current direction as a function of temperature. The samples were thinned down in two steps from 0.15 (i=1, curve (1)) to 0.09 mm (i=3, curve (3)). In the inset, an oriented sample with the four platinum wire contacts prepared for the measurement is shown.}
\label{bildrho}\label{LuNi4P2_Rho}\label{bildrho2}
\end{figure}

For a tetragonal material Ohm`s law reads
\begin{eqnarray}
\vec{j}=\left( \begin{array}{ccc}\sigma_a&&\\
                                    &\sigma_a&\\
                                    &&\sigma_c
\end{array}
\right)
\vec{E}, \quad\quad \rho = \frac{1}{\sigma}\label{Formel2}
\end{eqnarray}
with $\sigma_a$ beeing the isotropic conductivity 
in the basal plane and $\sigma_c$ the conductivity 
along the c-direction for the electrical field $\vec{E}$. 
The coordinates are choosen such that the z-axis is oriented 
along the crystallographic c-direction.
Our Czochralski grown YbNi$_4$P$_2$ single crystals reproduce 
the temperature dependence 
of the reported data as shown in the main part of Fig.~\ref{bildrho} 
and with them it is also possible to 
resolve the temperature dependence of the resistivity 
$\rho_{100}$ and $\rho_{110}$ of the in-plane directions. 
As expected according to Eqn.~\ref{Formel2}, 
the resistivity is isotropic in the a-a plane.
We determined the absolute values at room temperature 
applying 
\begin{eqnarray}
\rho(T) = \frac{a b}{l I}\, U(T)\label{Formel1}
\end{eqnarray}
with the width $a$, the thickness $b$ of the sample, 
the contact distance $l$ 
and the applied constant current $I$ and measured voltage $U$.
For $j\parallel 001$ we determined 
the resistivity $\rho_{001}(300\rm K)= 60\mu\Omega \rm cm$,
for $j\perp 001$ we found $\rho_{110}(300\rm K)= 70\mu\Omega \rm cm$.
The absolute value determined for the $[100]$-direction agrees 
with that of the $[110]$-direction.
For comparison, the resistivity of the non-magnetic reference LuNi$_4$P$_2$
was measured and is shown in the inset of Fig.~\ref{LuNi4P2_Rho}.
For the determination of the absolute values,  
the measurement device ($U$, $I$) as well as the geometry of the samples ($a$, $b$, $l$)
have to be considered as sources for uncertainties.
In the resistivity measurement, 
the main contribution to the uncertainty
comes from the geometry factors 
$\Delta a/a\simeq 3\%$, 
$\Delta b/b\simeq 3\%$ and $\Delta l/l\simeq 1\%$.
The  uncertainty of the voltage $\Delta U/U$
is between $1\%$ and $3\%$  for different  temperatures.
The uncertainty of the current \cite{ACTapplnote} 
with $\Delta I/I\simeq 5\cdot 10^{-4}\%$ is neglegible.

The aim of our optimization of the crystal growth 
process is to minimize the crystal defects. 
Indispensable is the use of high-purity starting materials. 
The contamination of the melt by crucible materials has to be avoided.
Low growth rates avoid flux inclusions.
Adjusting the temperature during the growth experiment 
reduces the evaporation of the elements
and therefore a shift in the stoichiometry.
At lowest temperatures, 
the scattering on crystal defects
is the only contribution to 
the resistivity.
Therefore, the resistivity ratio 
RR$_{1.8\rm K} = \rho(300 \rm K)/\rho(1.8\rm K)$ 
is an indicator for the amount of crystal defects. 

For 6 different samples, 
which were cut from the same single crystal, 
different  RR$_{1.8\rm K}^{001}$ values between
11 and 17 were found.
For the purest sample,
we determined the absolute resistivity value to be 
$\rho_{001}(1.8\rm K)=$ 3.5 $\mu\Omega \rm cm$.
In the other case,
if the current is perpendicular to the $c$-direction,
RR$_{1.8\rm K}^{110}$ and RR$_{1.8\rm K}^{100}$, respectively,
spread less for the 6 examined samples.
The values are between 3.0 and 3.07.
The absolute value is
$\rho_{110}(1.8\rm K)=$ 22.8 $\mu\Omega \rm cm$.
The large variation in RR$_{1.8\rm K}^{001}$ can 
be caused by a misalignment of the four contacts
for the current and voltage measurement.
During the growth
small-angle grain boundaries can evolve in a single crystal.
They lead to the distorsion of the lattice and  
enhance the residual resistivity. 
A further analysis by Laue method is required to clarify
the origin of the variation of RR$_{1.8\rm K}^{001}$.

A common technique to improve the crystallinity of materials 
is annealing. An experiment at 800$^\circ$ C in an inert atmosphere
for 5 days yields an increase of 25\% of RR$_{1.8\rm K}^{001}$. 
Due to the large anisotropy 
for the different current directions, 
it is not clear if this improvement is the result 
of the annealing procedure or 
was caused by a slight misalignment of the contacts during 
the resistivity measurement. 
For further annealing studies with different parameters,
well cut samples with a precise orientation are necessary. 
We investigated the influence of grinding and polishing
to the residual resistivity ratio.
Two samples cut from the single crystal were
oriented along the $[110]$ and the $[001]$-direction.
They were thinned down in two steps from 0.15 to 0.09 mm 
and the respective temperature dependence of the resistivity 
$\rho_1, \rho_2, \rho_3$ (curves (1-3)) was measured and is shown 
on the left hand side of Fig.~\ref{bildrho2}.
We determined from our data for each current direction 
the ratio $\rho_i/\rho_{\rm av}$ 
with $\rho_{\rm av}=(\rho_1\cdot\rho_2\cdot\rho_3)^{1/3}$ 
to investigate if our polishing procedure causes stress 
and strain in the samples and therefore influences 
the sample quality. 
The right part of Fig.~\ref{bildrho2}  
shows that RR$_{1.8\rm K}$ slightly varies between 
the different thicknesses labelled by (1), (2), (3). 
For the current parallel to the $c$-direction, the polishing diminishes
the value of $\rho_{001}(1.8\rm K)$.
For the perpendicular direction, the opposite effect occurs. 
This result might be caused by the anisotropy of RR$_{1.8\rm K}$. In the thinner sample
 the ideal current flow without a perpendicular component is approached. That means that 
the current is confined to the $[001]$ or the $[110]$-direction and the respective perpendicular current component is reduced upon thinning down the sample. For the $[001]$-direction $\rho_i$ becomes smaller for increasing $i$ leading to a larger RR$_{1.8\rm K}$ and for the $[110]$-direction $\rho_i$ becomes larger leading to a smaller RR$_{1.8\rm K}$ value. 
This investigation also demonstrates that the polishing procedure itself seems not to enlarge the number of crystal defects.


\section{The substituted compound YbNi$_4$(P$_{1-x}$As$_x$)$_2$, $x=0.2$}

\begin{figure}[ht]
\centering
\includegraphics[width=0.34\textwidth]{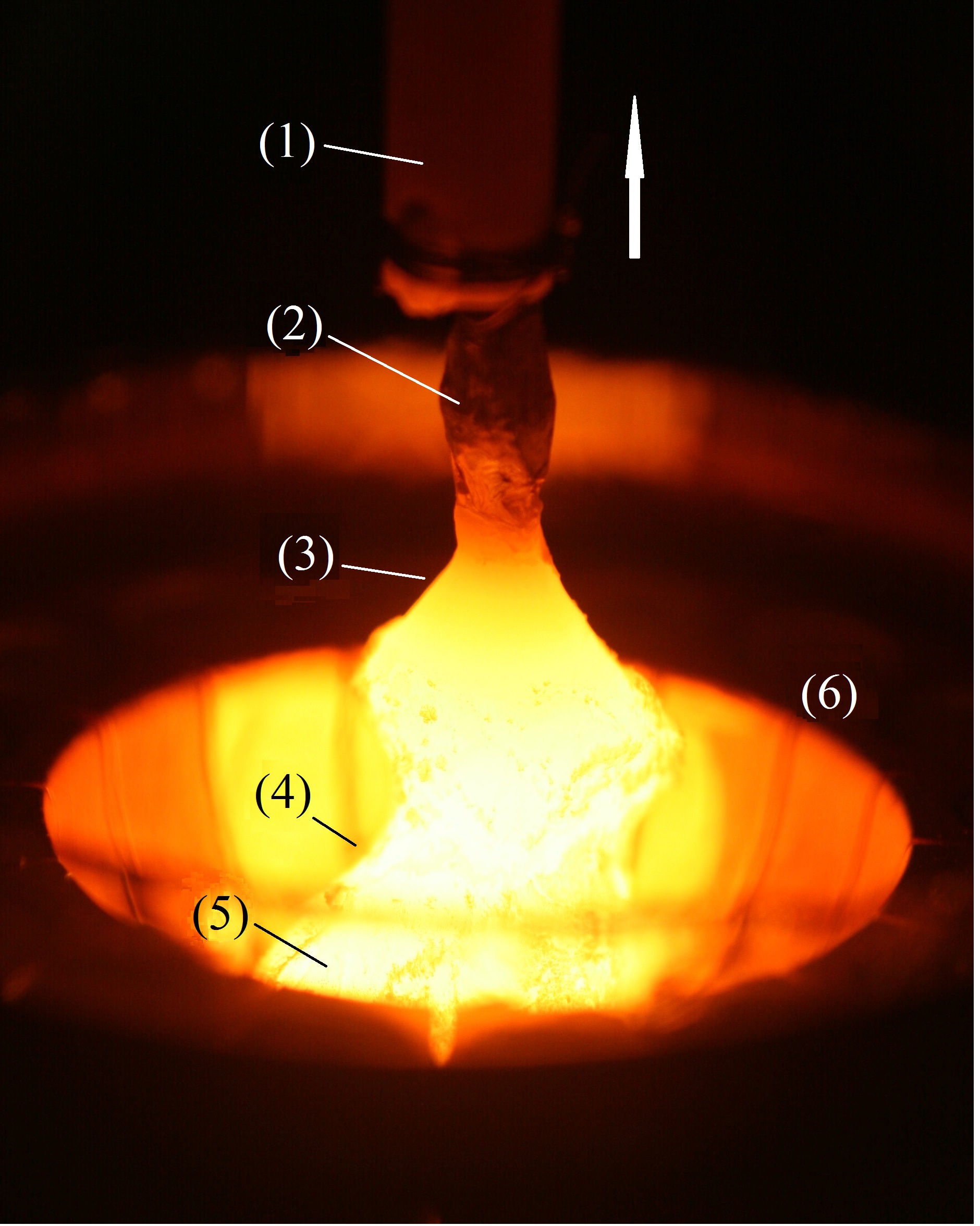}
\includegraphics[width=0.58\textwidth]{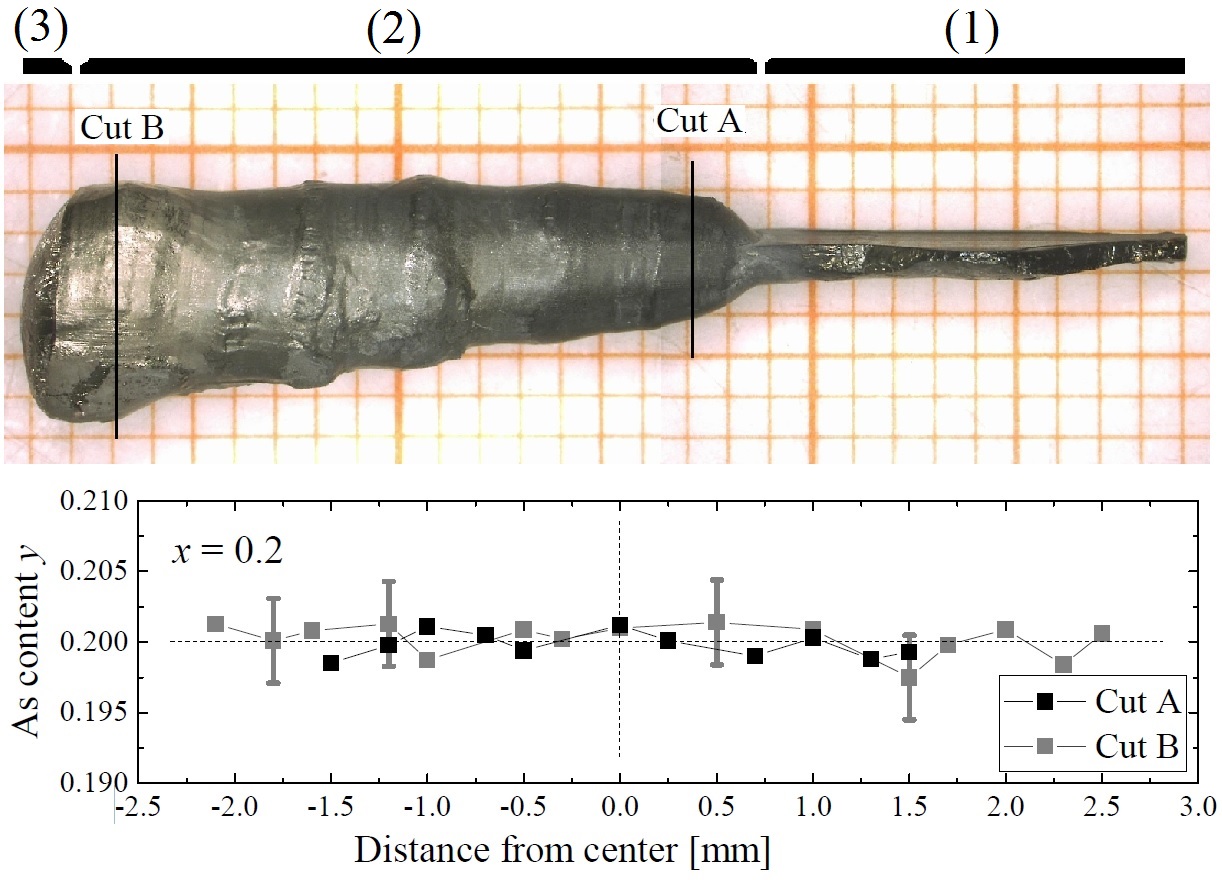}
\caption{{\it Left:} The YbNi$_4$(P$_{1-x}$As$_x$)$_2$, 
$x = 0, 0.2$ crystals were grown from a levitating melt. 
The arrow mark the pulling direction of the seed (2) in the seed holder (1). 
The meniscus (4) marks the border between the grown sample (3) 
and  melt (5) levitating in a cold copper crucible (6). 
The heating power is provided via a radio-frequency induction coil. 
The evaporation of the volatile elements was slowed down by applying 
an Ar pressure of 20 bar in the growth chamber;
{\it Right, upper part:} The as grown sample with the 
nominal composition of YbNi$_4$(P$_{0.8}$As$_{0.2}$)$_2$.  
The single crystal seed (1) consisted of YbNi$_4$P$_2$ 
and was oriented along the $[001]$-direction. 
The determination of the As content of region (2) 
of the grown sample is shown in the lower part of the figure. 
The growth was terminated with a faster growth velocity 
leading to the incorporation of flux in part (3) 
of the sample; 
{\it Right, lower part:} 
The real As content $y$ in the grown sample along 
two radial lines, A,B (black and grey symbols) 
was determined using EDX in dependence of the distance 
from the center of the sample.}
\label{043_Bild}
\end{figure}
The growth of the substituted 
compound by the Czochralski method Fig.~\ref{043_Bild}, 
was done similarily to that reported in \cite{Kliemt2016}.
The liquidus temperature is measured during the growth experiments.
Within the accuracy of the pyrometer $\Delta T \approx \pm 50 \rm K$, 
no change compared to the stoichiometric compound was detected.
The growth was started at about 1400$^{\circ}$C using an
YbNi$_4$P$_2$ single crystal as the seed.
After seeding, the sample was pulled with a rate of 0.3\,mm/h. 
Such low growth velocities are necessary
to obtain crystals without flux inclusions.
The distribution coefficient of As in the system 
$\kappa = c_l^{\rm As}/c_s^{\rm As}$  with the concentrations
of As in the melt $c_l^{\rm As}$ 
and in the solid $c_s^{\rm As}$, respectively, 
is not known and does not necessarily have to be exactly one \cite{Wilke1988}. 
A deviation from $\kappa = 1$ would lead to the enrichment 
or depletion of As in the melt and an inhomogeneous 
As distribution in the grown crystal.
The As distribution in the single crystal was investigated by EDX, 
and showed that the initial As:P ratio of the melt of 1:4 
can be found all over in our sample except at the first part 
which is connected to the seed.
 The seed consisted of YbNi$_4$P$_2$ and over a length 
of about 2.5 mm the crystal structure included more 
and more As and the lattice adapted to the new lattice constants.
 The analysis along radial lines (Cut A,B) on polished surfaces 
shown in Fig.~\ref{043_Bild}, yields a homogeneous As concentration 
from the center to the surface of the crystal.
A part of the YbNi$_4$(P$_{1-{\it x}}$As$_{\it x}$)$_2$, $x = 0.2$ 
sample was investigated by PXRD and the structure refinement 
yields an enlargement of the unit cell of $\Delta V/V=0.0138$ compared to $x = 0$. 
The lattice constants of the tetragonal $P4_2/mnm$ structure 
are $\rm a = 7.0803(5)\AA$ and $\rm c = 3.6131(2)\AA$.



\section{Summary}
The electrical resistivity was measured on Czochralski grown 
YbNi$_4$P$_2$ single crystals from 1.8~to~300\,K and
the absolute values at room temperature were determined to be 
$60\,\mu\Omega \rm cm$ for $j$ parallel to the $c$-direction and 
$70\,\mu\Omega \rm cm$ perpendicular. 
Comparing with the measurement on Bridgman grown samples 
\cite{Steppke2013},
the data reproduce the temperature dependence  of the resistivity 
but the absolute values differ.
The resistivity ratio RR$_{1.8 \rm K}$ is strongly anisotropic. 
For $j$ parallel to the $c$-direction
we found value between 11 and 17 for different samples.
In the perpendicular  directions, 
it is almost sample independent with values about 3.
Polycrystalline
LuNi$_4$P$_2$ samples, exhibit the resistivity characteristics
of a non-magnetic metal.
A single crystal of the substituted compound 
YbNi$_4$(P$_{1-x}$As$_x$)$_2$ with $x=0.2$ was grown.
The phosphorous to arsenic ratio 
in the melt is maintained  in the crystal. 
The As distribution is homogeneous throughout the whole sample.
The enlargement of the unit cell is $\Delta V/V=0.0138$ compared to $x = 0$.


\section*{Acknowledgments}
We thank W. Assmus and F. Ritter for valuable discussions 
and K.-D. Luther for technical support. The authors grateful acknowledge support by the DFG
via project KR 3831/4-1.


\bibliography{YbNi4P2_ConfSer_resubmitted_final_arxive.bbl}

\begin{thebibliography}{10}

\bibitem{Krellner2011}
C.~Krellner, S.~Lausberg, A.~Steppke, M.~Brando, L.~Pedrero, H.~Pfau,
  S.~Tenc{\'e}, H.~Rosner, F.~Steglich, and C.~Geibel.
\newblock {\em New J. Phys.}, 13:103014, 2011.

\bibitem{Brando2016}
M.~Brando, D.~Belitz, F.M. Grosche, and T.R. Kirkpatrick.
\newblock {\em Rev. Mod. Phys.}, 2016 (accepted).

\bibitem{Steppke2013}
A.~Steppke, R.~K\"uchler, S.~Lausberg, E.~Lengyel, L.~Steinke, R.~Borth,
  T.~L\"uhmann, C.~Krellner, M.~Nicklas, C.~Geibel, F.~Steglich, and M.~Brando.
\newblock {\em {Science}}, 339:933, 2013.

\bibitem{Krellner2012}
C.~Krellner and C.~Geibel.
\newblock {\em J. Phys. Conf. Ser.}, 391:012032, 2012.

\bibitem{Kliemt2016}
K.~Kliemt and C.~Krellner.
\newblock {\em J. Cryst. Growth}, 449:129 -- 133, 2016.

\bibitem{Chikhrij1986}
S.~I. Chikhrij, S.~V. Orishchin, and Y.~B. Kuz$^{\prime}$ma.
\newblock {\em Dopov Akad Nauk A}, 7:79, 1986.

\bibitem{Jeitschko1990}
W.~Jeitschko, F.~Terb\"uchte, E.J. Reinbold, P.G. Pollmeier, and T.~Vomhof.
\newblock {\em J. Less. Common Met.}, 161:125, 1990.

\bibitem{GSAS}
A.C. Larson and R.B.~Von Dreele.
\newblock {\em Los Alamos National Laboratory Report LAUR}, page~86, 2000.

\bibitem{EXPGUI}
B.~H. Toby.
\newblock {\em J. Appl. Crystallogr.}, 34:210, 2001.

\bibitem{ACTapplnote}
QuantumDesign.
\newblock {\em {Electro-transport Application note, Physical Property
  Measurement System}}, 2016.

\bibitem{Wilke1988}
K.-T. Wilke and J.~Bohm.
\newblock {\em Kristallz\"uchtung}.
\newblock VEB Deutscher Verlag der Wissenschaften, Berlin, 1988.

\end{thebibliography}
\end{document}